\documentclass[twocolumn,aps,floatfix,superscriptaddress]{revtex4-2}

\usepackage{graphicx,color}
\usepackage{dcolumn}
\usepackage{epsfig}
\usepackage{changes}
\usepackage[centertags]{amsmath}
\usepackage{amsfonts,amsmath}
\usepackage{euscript}
\usepackage{amssymb}
\usepackage{amsthm}
\usepackage{newlfont}
\usepackage{mathrsfs}
\usepackage{subfigure}
\usepackage{hyperref}
\usepackage{mathptmx}

\newcommand{\opunit}{\text{1}\kern-0.22em\text{l}}

\newcommand{\id}{\textrm{d}}

\def\bea{\begin{eqnarray}}
\def\eea{\end{eqnarray}}
\def\ba{\begin{array}}
\def\ea{\end{array}}

\begin{document}

\title{
Ergodicity Breaking in Active Run-and-Tumble Particles in a Double-Well Potential
} 
\author{Urna Basu} 
\affiliation{S. N. Bose National Centre for Basic Sciences,  Kolkata 700106,  India}
\author{Satya N. Majumdar}  
\author{Alberto Rosso}
\affiliation{LPTMS,  CNRS,  Universit{\'e} Paris-Saclay,  91405 Orsay,  France}

\begin{abstract}
We investigate the dynamics of a run-and-tumble particle in a double-well potential and demonstrate that, in stark contrast to Brownian particles, active dynamics can lead to strong ergodicity breaking. When the barrier height exceeds a critical threshold, the long-time position distribution depends crucially on the initial condition: if the particle starts within the basin of attraction of one well, it remains trapped there, while if it begins between the two basins, it can reach either well with a finite probability, which we compute exactly via hitting probabilities. Below the critical barrier height, ergodicity is restored and the system converges to a unique stationary distribution, which we derive analytically. Using this result, we also estimate the characteristic barrier crossing time and show that it violates Kramer's-Arrhenius law, and displays a divergence near the critical height  following a Vogel-Fulcher-Tammann-like form with an anomalous exponent $1/2$.
\end{abstract}

 
\maketitle

Active particles -- from synthetic Janus colloids to motile cells -- display striking behaviors with no counterpart in equilibrium systems~\cite{romanczuk2012,RevModPhys.88.045006,o2022time,fodor2018statistical}. Their persistent motion gives rise to qualitatively new phenomena at the collective level, such as flocking~\cite{Giardina2014}, motility-induced phase separation~\cite{cates2015motility}, and confinement-induced effects~\cite{caprini2018active, caprini2021collective} as well as at the single-particle level, including anomalous position fluctuations~\cite{Basu2018,santra2021active,1drtp,GradenigoMajumdar2019,MoriGradenigoMajumdar2021,SmithFarago2022,Smith2023} and unusual first passage properties~\cite{mori2020universal,MoriLeDoussalMajumdarSchehr2020,nath2024survival, gueneau2025run}. In the absence of thermal fluctuations, a single particle in an external confining potential shows a non-Boltzmann stationary state, being typically bounded in a finite region, with its position most likely to be a finite distance away from the minimum of the potential in the limit of strong activity~\cite{dhar2019run,PhysRevE.100.062116,santra2021direction,chaudhuri2021active,smith2022exact}. This raises an important question about the position fluctuations of active particles in landscapes with multiple minima separated by finite barriers, which is virtually unexplored.

A related question is how the escape time of a particle across a barrier is affected by the presence of activity. For passive Brownian particles, the typical time to cross a barrier of height $\Delta V$ is given by the Kramer's law
$\tau \sim \exp(\beta \Delta V)$, where $\beta$ is the inverse temperature~\cite{kramers1940brownian,rondin2017direct}. However, how much of this universal feature remains valid in the presence of activity is an open question. This question has 
been investigated in the presence of single or multiple barriers, in the limits of strong and weak activity~\cite{Sharma2017,geiseler2016kramers,woillez2019activated,caprini2019active,wexler2020dynamics}. Recently, the transition rate between the two wells for an active particle in a double well potential has also been measured experimentally~\cite{militaru2021escape}, which shows the existence of an optimum activity which maximizes the transition rate. However, most of these works focus on the finite temperature scenario, where thermal noise already allows the particle to cross the barrier. How the pure active nature of the dynamics, without additional thermal fluctuations, affects the barrier crossing time however remains elusive and is one of the
main focuses of this work. 


\begin{figure}[t]
 \centering
 \includegraphics[width=7.2 cm]{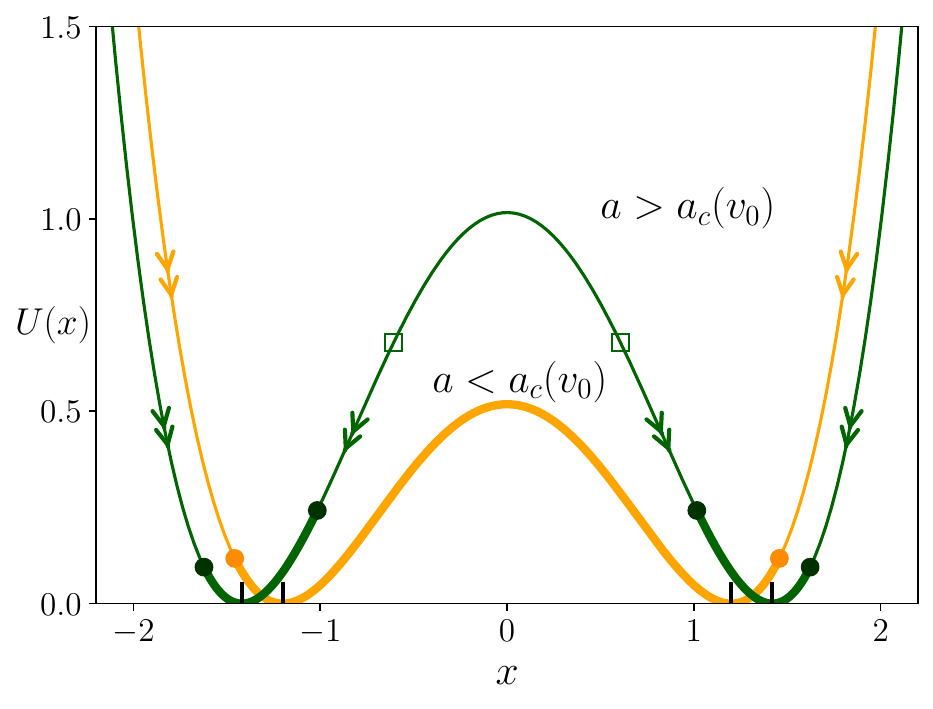}
 \caption{Schematic representation of the double well potential for $a<a_c$ (light orange curve)  and $a>a_c$ (dark green curve). The thick sections of the curves indicate the support of the stationary position distribution in the respective cases, and the arrows indicate the direction of the net force on the RTP. For $a> a_c$ the empty squares indicate the points $x=\pm \ell_1$ and the filled discs indicate $x=\pm \ell_2$ and $x=\pm \ell_3$.}\label{fig:pot}
\end{figure}





In this Letter, we consider the simplest non-trivial setting, namely, motion of an active Run-and-Tumble Particle (RTP) in a double-well potential in one-dimension. We consider a symmetric double well potential $U(x) = \frac 14 (x^2 -a^2)^2$ which has two minima at $x = \pm a$ separated by a barrier of height $U(0) = a^4/4$ [see Fig \ref{fig:pot}]. Such potentials are paradigmatic set-up for investigation of barrier crossing phenomenon.  For a passive Brownian particle initially localized in one well, the probability of crossing to the other well is governed by the Kramer's/ Arrhenius law which leads to a finite mean escape time. Consequently, the particle eventually explores both wells and relaxes to a unique stationary Boltzmann distribution, a result that holds irrespective of the precise shape of the potential. In this work we show that the picture 
is fundamentally different for an active particle, both for position fluctuations and barrier crossing time. We begin by analyzing the stationary position distribution of an RTP in a double-well potential.

The position $x(t)$ of an RTP in the double-well potential $U(x)$  evolves following the Langevin equation,
\bea
\dot x(t) = f(x) + v_0 \sigma(t). \label{eq:langevin} 
\eea
where,
\bea
f(x) \equiv - U'(x) = - x(x^2-a^2), \label{eq:fx}
\eea
is the force on the particle from the potential. In the absence of any external force, the particle moves with a constant speed $v_0$ along its internal orientation $\sigma$,  which alternates between $\pm 1$ with a rate $\gamma$. 

\begin{figure}[t]
 \centering
 \includegraphics[width=8 cm]{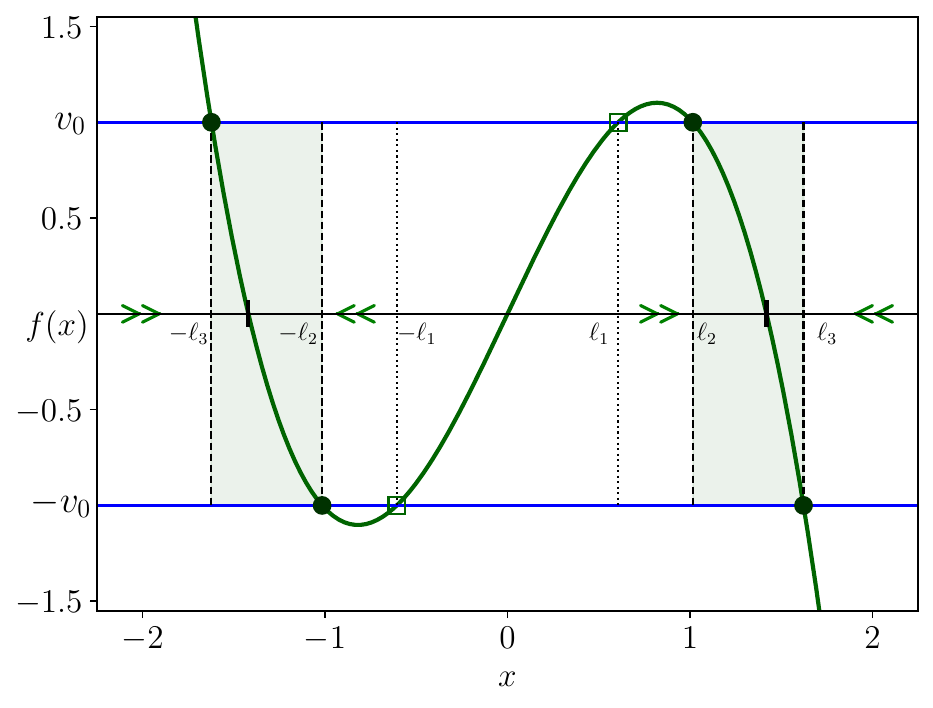}
 \caption{Schematic representation of the fixed points of the force $f(x)$ for $a>a_c$. Here we have taken $a=1.42$ and $v_0=1$ so that $a_c\simeq 1.3747$. The shaded green regions indicate the supports of the stationary distribution. The dotted lines indicate the boundaries of the basins of attractions of the two wells. The arrows along the $x$-axis indicate the direction of the force on the particle in the respective regions.}\label{fig:force}
\end{figure}

The bounded nature of the active dichotomous noise $\sigma(t)$ implies that the RTP cannot dynamically access the regions where $|f(x)| > v_0$. Thus, the turning points, where $f(x) = \pm v_0$, act as effective walls, confining the particles in a finite region. 
For the double-well potential,  the force $f(x)$ is a non-monotonic and odd function of $x$,  with extrema at $x=\pm m$ where $m=a/\sqrt{3}$ [see Fig.~\ref{fig:force}].  Clearly,  if $ |f(m)| = 2 a^3/ 3 \sqrt{3} < v_0$, there is a unique solution for $f(x)=v_0$. Thus, for $a < a_c$, where,
\begin{align}
a_c  = \left(\frac{3 \sqrt{3}\, v_0}{2}\right)^{1/3}, \label{eq:ac_def}
\end{align}
there are only two turning points $x= \pm \ell (a)$ and the particle will be confined in the region $ -\ell(a) \le  x \le \ell(a)$ [see Appendix for more details]. The corresponding stationary distribution is ergodic, and can be obtained using the generic result derived in Refs.~\cite{solon2015pressure, dhar2019run} for arbitrary confining potentials,
\bea
P_\text{st}(x)= \frac{\cal N}{v_0^2-f^2(x)} \exp{\bigg[2 \gamma \int_0^x \frac{\id u~ f(u)}{v_0^2 - f^2(u)}\bigg]}, \label{eq:pst_gen}
\eea
where $\cal N$ is the normalization constant, given by,
\bea
\cal N^{-1} = \int_{-\ell}^{\ell} \frac{\id x}{v_0^2 - f^2(x)} \exp{\bigg[2 \gamma \int_0^x \frac{\id u~ f(u)}{v_0^2 - f^2(u)}\bigg]}. \label{eq:norm}
\eea

The scenario becomes much more complex and interesting for $a > a_c$, where the equation $f(x) = v_0$ has three real solutions. Consequently, there are six turning points, denoted $\pm \ell_{1}$, $\pm \ell_{2}$, and $\pm \ell_{3}$; for notational convenience we assume $\ell_3 > \ell_2 > \ell_1 >0$. These turning points divide the phase space (the real line) into seven distinct regimes, as illustrated in Fig.~\ref{fig:force}. 
The splitting of phase space into disconnected regions results in a position distribution that is inherently non-ergodic. In the following we first investigate the position fluctuations in the 
non-ergodic regime $a > a_c$.\\


\begin{figure}[tb]
 \centering
 \includegraphics[width=8.2 cm]{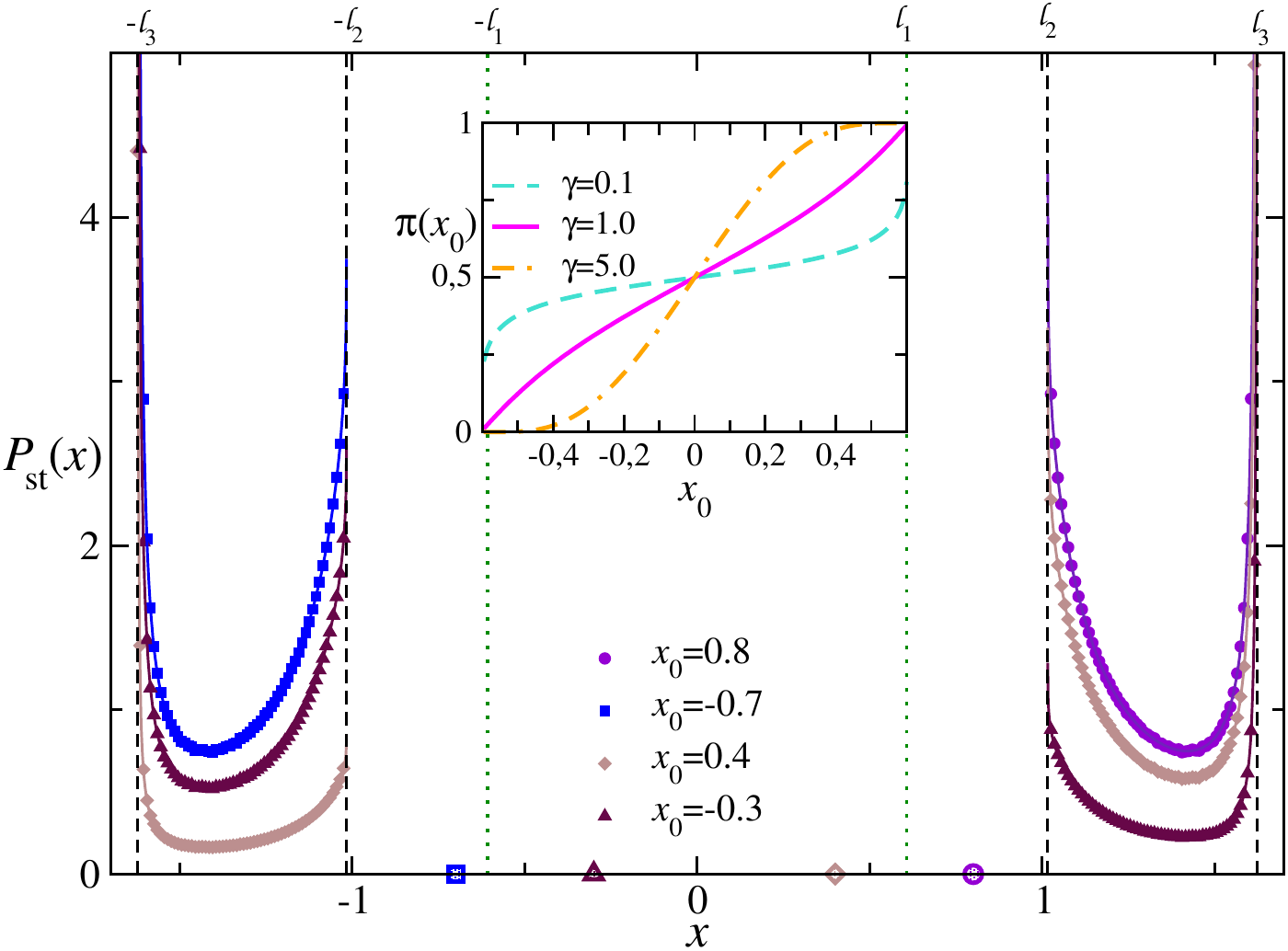}
 \caption{Non-ergodic phase: 
  Plot of $P_\text{st}(x)$ for $a=1.42> a_c$ and different values of initial position $x_0,$ for $\gamma=1$. The solid lines correspond to the analytical prediction Eq.~\eqref{eq:P2w} and symbols to numerical simulations. The dotted vertical lines indicate $x=\pm \ell_1$ whereas the dashed vertical lines indicate $x= \pm \ell_2$ and $x=\pm \ell_3$. Here we have taken $v_0=1$ so that $a_c \approx 1.3747$ and $\ell_1 \approx 0.6067, \ell_2 \approx 1.0159, \ell_3 \approx 1.6226$. The symbols on the $x$-axis mark the values of $x_0$ corresponding to the four curves. The inset shows the plot of the hitting probability $\pi(x_0)$ versus $x_0$ in the regime $-\ell_1 \le x_0 \le x_0$ for different values of $\gamma$. The values of $a$ and $v_0$ used are same as in the main plot. }
 \label{fig:Px_nonergo}
\end{figure}

\noindent {\it Non-ergodic regime}: To understand the behaviour of the stationary position distribution for $a> a_c$,  it is useful to consider the Langevin equation \eqref{eq:langevin} and investigate the net force acting on the particle in the different regions.  
For $x> \ell_3$ and $\ell_1 \le x \le \ell_2$, $|f(x)| > v_0$, and the particle feels an attractive force towards the right well, irrespective of its internal orientation [see Fig.~\ref{fig:force}]. Furthermore, once the particle enters the region $\ell_2 \le x \le \ell_3$ it cannot escape. Thus, the region  $x > \ell_1$ forms the `basin of attraction' for the right well---if the initial position of the particle $x_0> \ell_1$, the particle eventually falls in the right well and gets trapped in the region $\ell_2 \le x \le \ell_3$.  Similarly, if initially the particle is in the region $x < -\ell_1$, it eventually falls in the left well, and remains trapped in the region  $-\ell_3 \le x \le -\ell_2$. However, if the initial position of the particle is in the intermediate domain,  i.e., $|x_0| < \ell_1$, then the direction of the net force on the particle depends on its orientation $\sigma(t)$, and it can move in both directions until getting trapped in one of the wells. Thus, for $a> a_c$, i.e., when the barrier height is larger than a critical value, the position distribution of the RTP depends on its initial position, leading to a breakdown of ergodicity in the stationary state. Such ergodicity breaking is a direct consequence of the active nature of the particle, and are absent in passive (Brownian) systems. In the following, we characterize this non-ergodic position distribution, which constitutes one of the central results of this work.

First, we consider the case $x_0 > \ell_1$, i.e., when the particle starts within the basin of attraction of the right well. As mentioned above, in this case, the particle is confined in the region  $\ell_2 \le x \le \ell_3$ in the steady state. The corresponding stationary distribution can be obtained using Eq.~\eqref{eq:pst_gen}, 
\bea
P_R(x) = \frac{\cal N_R}{v_0^2 - f^2(x)} \exp{ \bigg[2 \gamma \int_{a}^x \frac{\id u~ f(u)}{v_0^2 - f^2(u)}\bigg]}.\label{eq:PL}
\eea
Note that, this distribution is supported in the region $\ell_2 \le x \le \ell_3$  and the normalization constant is given by,
\bea
\cal N_R^{-1} = \int_{\ell_2}^{\ell_3} \frac{\id x}{v_0^2 - f^2(x)}\exp{ \bigg[2 \gamma \int_{a}^x \frac{\id u~ f(u)}{v_0^2 - f^2(u)}\bigg]}. \label{eq:NL}
\eea

Next, we consider the case $x_0 < -\ell_1$, i.e., when the particle starts within the basin of attraction of the left well. The symmetry of the potential implies that the stationary distribution of the RTP in this case is given by,
\begin{align}
P_L(x) = P_R(-x) .
\end{align}
Obviously, $P_L(x)$ is supported in the region $-\ell_3 \le x \le -\ell_2$. Figure~\ref{fig:Px_nonergo} shows plots of the position distribution of the RTP when it is trapped in one of the wells. 

\begin{figure}[t]
 \centering
 \includegraphics[width=7.25 cm]{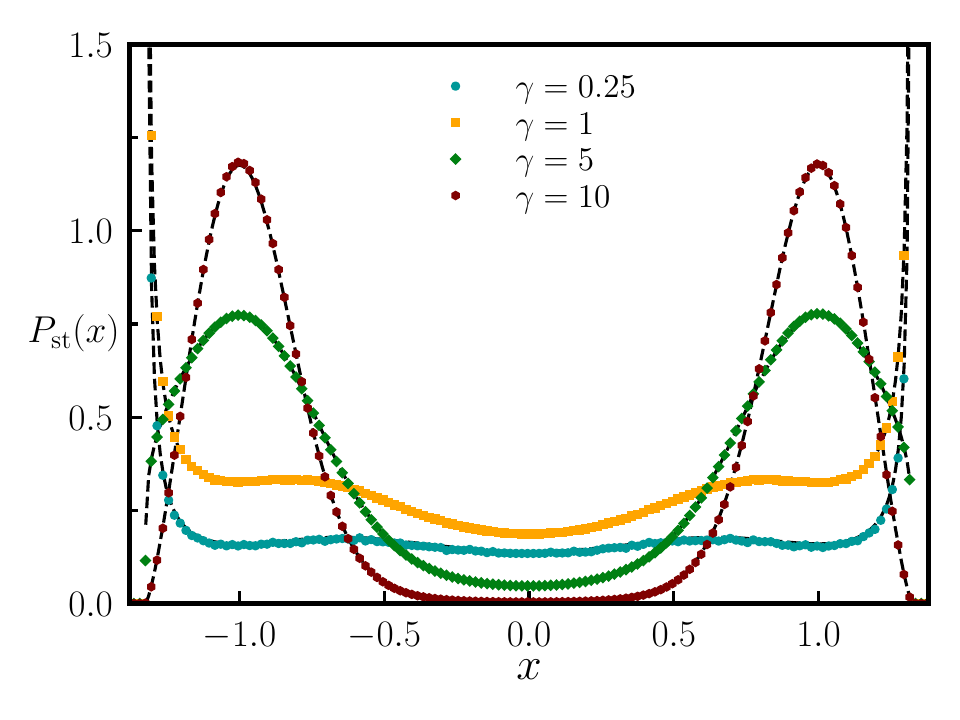}
 \caption{Position distribution in the ergodic regime: Plot of $P_\text{st}(x)$ {\it vs} $x$ for $a=1$ for different values of $\gamma.$ The symbols correspond to the data obtained from numerical simulations while the dashed black lines correspond to the analytical prediction Eq.~\eqref{eq:pst_gen}. }\label{fig:a1}
\end{figure}

The most non-trivial situation emerges when the initial position $x_0$ falls outside the basins of attraction of both the wells,  i.e.,  in the intermediate domain $|x_0| < \ell_1$. In this case, the RTP moves randomly until it reaches one of the basins, and then gets trapped in the corresponding well. Thus, the position distribution depends explicitly on the initial position when it lies in the intermediate region between the basins of attraction of the two wells. 
In this case, we can define the position distribution $P(x;x_0)$ averaged over all possible realizations with a fixed initial position $x_0$. To compute it,  we need to find the probability of the RTP exiting the region $-\ell_1 \le x \le \ell_1$ through the right or left boundary. This is the classic `hitting probability', sometimes also referred to as `splitting probability'---for a one-dimensional stochastic process $x_t$ in an interval, the 
hitting probability quantifies the likelihood that starting at a fixed position $x_0$, the process reaches one boundary of the interval before the other. For ordinary Brownian motion this probability is known, both in free space~\cite{redner2001guide} and in presence of an arbitrary potential~\cite{majumdar2010hitting}. The hitting probability for an RTP particle in an arbitrary potential has also been computed very recently~\cite{gueneau2024relating} which we rederive in the Appendix \ref{app:hitting prob} using our notations. Our results are summarized as follows.

Let $Q_\sigma(x_0)$ denote the probability that, starting from $x_0$ with internal orientation $\sigma$, the RTP hits the right boundary $x=\ell_1$ before hitting the left boundary $x=-\ell_1$. It can be shown that , $Q_\sigma(x_0)$ satisfies the backward Fokker-Planck equation,
\begin{align}
[f(x_0) + \sigma v_0] \frac{d Q_\sigma}{d x_0} + \gamma[Q_{-\sigma}(x_0) - Q_\sigma(x_0)] =0,
\end{align}
with the boundary conditions $Q_+(\ell_1)=1$ and $Q_-(-\ell_1)=0$ [see Appendix \ref{app:hitting prob} for details]. This set of coupled first order differential equations can be solved exactly,
to obtain,
\begin{align}
Q_+(x_0) &= \frac{1}{1+\gamma G} \left[ 1+ \gamma \int_{-\ell_1}^{x_0} \frac{\id y}{v_0 + f(y)} e^{\textstyle -2 \gamma \int_{-\ell_1}^y \frac{\id u~ f(u)}{v_0^2 - f^2(u)} } \right],\cr
Q_-(x_0) &= \frac{\gamma}{1+\gamma G}  \int_{-\ell_1}^{x_0}\frac{\id y}{v_0 - f(y)} e^{\textstyle -2 \gamma \int_{-\ell_1}^y \frac{\id u~ f(u)}{v_0^2 - f^2(u)} }, \label{eq:QpQm}
\end{align}
where, 
\bea
G = \int_{-\ell_1}^{\ell_1} \frac{\id y}{v_0+f(y)} \exp{\bigg[-2 \gamma \int_{-\ell_1}^y \frac{\id u~ f(u)}{v_0^2 - f^2(u)}\bigg]}.\,
\eea
Thus, the probability that the RTP,  starting from the intermediate region $-\ell_1 \le x_0 \le \ell_1$, gets trapped in the right well, is given by,
\bea
\pi(x_0) = \frac 12 \bigg[Q_+(x_0)+Q_-(x_0)\bigg],
\eea
where we have assumed that the initial orientation $\sigma=\pm 1$ with equal probability $1/2$.  Inset of Fig.~\ref{fig:Px_nonergo} shows plot of $\pi(x_0)$ as a function of $x_0$.  Clearly,  it increases monotonically indicating the particle preferably exits the interval via the closer boundary. However, the non-linear nature is a signature of the active nature of the motion.  

Using the above results, we arrive at the average position distribution in the long-time limit, 
\bea
P(x; x_0) = \pi(x_0) P_\text{st}^R(x) + [1- \pi(x_0)] P_\text{st}^L(x). \label{eq:P2w}
\eea
 Figure~\ref{fig:Px_nonergo} compares the above predicted distribution with the same obtained from numerical simulations for two different values of $x_0$, which show excellent agreement. Note that, in this case, the stationary distribution cannot be obtained by time-averaging and one has to average over independent trajectories to measure the distribution correctly. This is an example of strong non-ergodicity, which has also been observed in the context of RTP in a periodic potential~\cite{le2020velocity}. Ergodicity breaking in a weaker sense has been also observed in the context of anomalous diffusion processes~\cite{cherstvy2015ergodicity,metzler2014anomalous,cherstvy2013anomalous}. \\
 


\noindent{\it Ergodic regime}: It is also useful to look at the shape of the stationary position distribution in the ergodic phase, i.e., for $a < a_c$.  In the ergodic phase the stationary distribution is unique, and is given by Eqs.~\eqref{eq:pst_gen}-\eqref{eq:norm}. As mentioned before, in this case, the distribution is supported in the regime $-\ell \le x \le \ell$. Although, it is hard to obtain a closed form for the position distribution, it can be estimated with arbitrary accuracy by performing the integral numerically. Figure~\ref{fig:a1} illustrates how the shape of $P_\text{st}(x)$ changes with the flip rate $\gamma$--the position distribution undergoes a shape transition as $\gamma$ is varied. Similar to an RTP in a harmonic potential~\cite{dhar2019run}, the distribution is strongly peaked near the boundaries $x= \pm \ell$ for small $\gamma$. For large $\gamma$, the particle effectively behaves like a Brownian one, leading to a Boltzmann-like distribution with two peaks around $x=\pm a$. \\

\begin{figure}[t]
 \centering
 \includegraphics[width=8.7 cm]{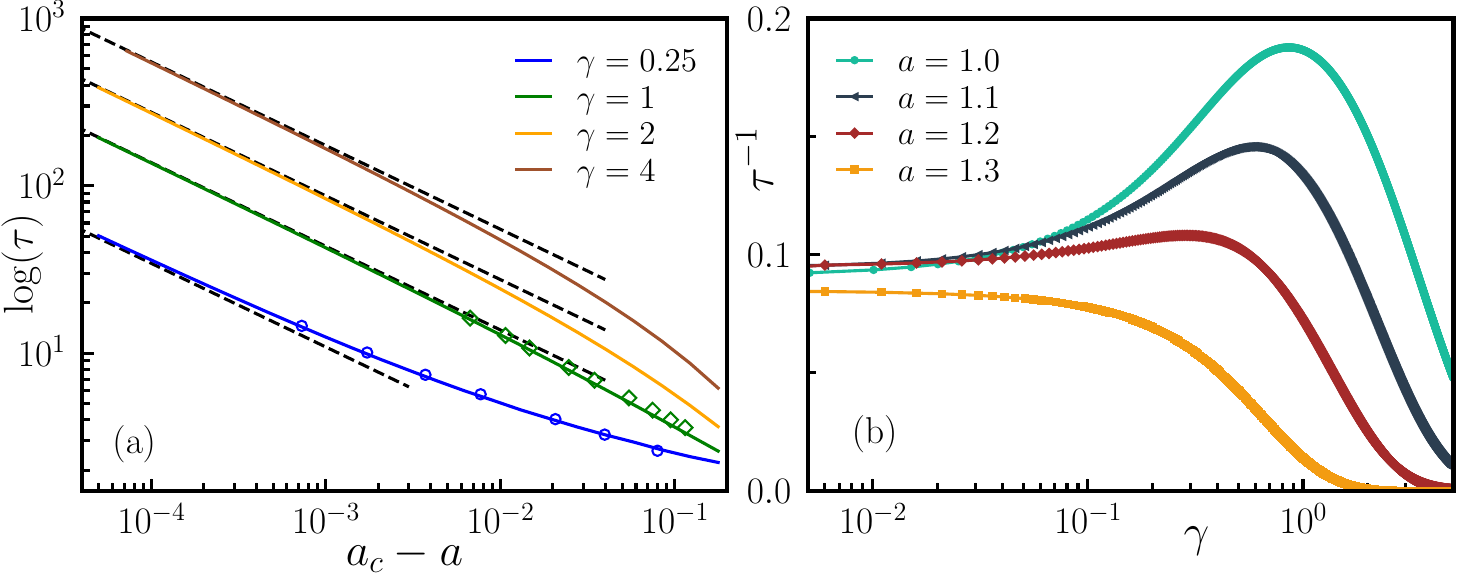}
 \caption{Barrier-crossing time in the ergodic phase: (a) Plot of $\log(\tau)$ versus $a_c-a$ in the log-log scale for different values of flip-rate $\gamma$. The solid lines indicate the data obtained from numerically integrating Eq.~\eqref{eq:norm} while the dashed black lines indicate the near-critical ${(a_c -a)}^{-1/2}$ behaviour predicted in Eq.~\eqref{eq:log_tau}. The symbols indicate the characteristic dwelling time measured from simulations. (b) Plot of escape rate as function of $\gamma$ for different values of $a$, obtained from numerically integrating Eq.~\eqref{eq:norm} and using Eq.~\eqref{eq:tau_def}.}
 \label{fig:escape}
\end{figure}


\noindent{\it Barrier crossing:} The double-well set-up provides an ideal playground to study the barrier crossing phenomenon for active particles. The most important quantity of interest in this regard is the characteristic timescale associated with crossing the barrier between the two wells. In the non-ergodic phase, the particle remains localized in one of the wells indefinitely, and therefore the barrier-crossing time diverges. In contrast, in the ergodic phase $a< a_c$, the RTP is able to cross the barrier and the corresponding time remains finite. A convenient measure of this timescale can be constructed from the weight of the stationary distribution at the location of the barrier. More precisely, we define,
\begin{align}
    \tau = 1/[v_0 P_\text{st}(0)] = v_0 {\cal N}^{-1}, \label{eq:tau_def} 
\end{align}
where ${\cal N}$ is given in Eqs.~\eqref{eq:pst_gen}-\eqref{eq:norm}. This, in turn, enables an analytical study of the barrier-crossing time, allowing us to investigate how $\tau$ scales as the barrier height approaches the critical threshold $a_c$.

To analyze the behaviour of ${\cal N}$, let us first note that, for $a=a_c$, we have $|f(x)|=v_0$ at $x=\pm m$. Consequently, the integrand in Eq.~\eqref{eq:norm} diverges as $a \to a_c$ and for small $\epsilon=a_c -a$, the dominant contribution to the integral comes from near $x=m$. This observation allows us to compute the asymptotic behaviour of $P_\text{st}(0)$ for small $\epsilon$, which finally leads to [see Appendix \ref{ap:time-scale} for the details], 
\begin{align}
\tau \propto \frac{1}{\sqrt{v_0(a_c -a)}} \exp \left[\frac{\pi \gamma}{3^{3/4} \sqrt{v_0 (a_c - a)} } \right]. \label{eq:tau_a}
\end{align}
Consequently, as $a \to a_c$, the leading asymptotic behaviour of the barrier-crossing time becomes,
\begin{align}
    \log \tau \approx \frac{\pi \gamma}{3^{3/4}} [v_0 (a_c - a)]^{-1/2}. \label{eq:log_tau}
\end{align}
Figure~\ref{fig:escape}(a) illustrates this algebraic divergence of $\log \tau$. 

Another measure of the barrier-crossing time is provided by the decay time scale of the dwelling-time distribution of the RTP within a well, which typically exhibits an exponential tail. This characteristic decay time, measured from numerical simulations, shows excellent agreement with the analytically predicted $\tau$, as illustrated in Fig.~\ref{fig:escape}(a).

The behaviour of the barrier crossing time, given in Eq.~\eqref{eq:log_tau} constitutes the second key finding of this work. It shows that for an RTP, the barrier crossing phenomenon is fundamentally different than that for the passive particles--without the help of thermal noise, the active particle is able to cross a barrier only when its height is smaller than a threshold value, which depends on its self-propulsion velocity. Moreover, the typical time required to cross the barrier shows a super-Arrhenius growth, diverging as the height approaches the threshold value. The divergence resembles a Vogel–Fulcher–Tammann law seen in glassy systems~\cite{garca1989theoretical}, albeit with an anomalous exponent $1/2$ instead of the standard exponent $1$.

It should be emphasized that the crucial  element underlying this strong ergodicity breaking and divergence of barrier crossing time for an RTP is the bounded nature of the dichotomous active noise $\sigma(t)$. In contrast, the active Ornstein–Uhlenbeck process (AOUP)~\cite{martin2021statistical}, driven by Gaussian noise, is not expected to exhibit strong ergodicity breaking. Indeed, for an AOUP in a double-well potential, the barrier-crossing time displays an Arrhenius-like dependence with an effective barrier height in the weak-noise limit~\cite{bray1989instanton}.


A recent experiment on active Brownian particles in a double-well potential~\cite{militaru2021escape} reported that the transition rate between the wells, equivalent to $\tau^{-1}$, exhibits a pronounced non-monotonic dependence on the rotational diffusion constant, reaching a maximum at an optimal value. In this context, it is worthwhile to investigate how $\tau$ depends on the flip rate $\gamma$, especially for $a \ll a_c$,  where Eq.~\eqref{eq:log_tau} is not expected to hold.  In this regime, $\tau$ can be obtained by numerically evaluating the integral in Eq.~\eqref{eq:norm}, which reproduces the experimentally observed nonmonotonic behavior. As shown in Fig.~\ref{fig:escape}(b), the escape rate $\tau^{-1}$ varies nonmonotonically with $\gamma$, attaining a maximum at an intermediate flip rate that shifts to smaller values as $a \to a_c$. The physical origin of the non-monotonicity can be understood as follows: For small $\gamma$, the particle is most likely to be near $\pm \ell(a)$, i.e., away from the barrier, which makes it harder for it to cross the barrier which leads to a large value of  $\tau$. As $\gamma$ is increased, the particle has a larger chance to be near the barrier [cf. the $\gamma=1$ curve in Fig.~\ref{fig:a1}], thus, reducing the $\tau$. For very large $\gamma$, the RTP typically behaves like a passive particle, with the position distribution being peaked at $\pm a$, i.e., it goes farther away from the barrier, thus increasing $\tau$ again. \\

\noindent {\it Conclusions:}  We have shown that a run-and-tumble particle in a double-well potential exhibits a sharp ergodicity-breaking transition controlled by the barrier height. Above a critical threshold, the dynamics partition the phase space into disconnected regions, leading to stationary distributions that depend on the initial condition; below it, ergodicity is restored and the unique stationary state can be obtained exactly. Our analysis further shows that an active particle behaves in a fundamentally different manner from its passive counterpart in barrier-crossing processes, with a strong violation of the Arrhenius' law.  Our results are easily generalized for multi-well situations and can possibly  be  verified  experimentally using a set-up similar to Ref.~\cite{militaru2021escape}.  It would be interesting to see if the ergodicity breaking transition
found here also show up for interacting many particle systems. A promising recent result in that direction was found in Ref.~\cite{touzo2025non}, where similar breaking of ergodicity was observed in a one-dimensional system of RTPs interacting via pair-wise double-well potential.

\acknowledgements
The authors thank Gregory Schehr for useful discussions. UB acknowledges the support from the Anusandhan National Research Foundation (ANRF), India, under a MATRICS grant [No.~MTR/2023/000392]. SNM acknowledges support from ANR Grant No.~ANR-23-CE30-0020-01 EDIPS.




\appendix




\section{Computation of the hitting probability}\label{app:hitting prob}

In this section, we outline the steps leading to the explicit computation of the hitting probability $\pi(x_0)$.  Let $Q_\sigma(x_0)$ denote the probability that, starting from some initial position $x_0$ with orientation $\sigma$,  the 
RTP exits the intermediate region through the right boundary $x=\ell_1$. To compute this probability, it is most convenient to adopt the backward Fokker-Planck equation approach.  Considering the possible trajectories over the interval $[0,\Delta t]$, we can write the backward FP equation,
\bea
Q_\sigma(x_0) = (1-\gamma \Delta t) Q_\sigma (x_0 + (f(x_0) + v_0 \sigma) \Delta t ) \cr
 + \gamma \Delta t Q_{-\sigma}(x_0),
\eea
where the first term on the right hand side corresponds to the 
contribution from the trajectories with no flip during this interval and the second term denotes the contribution from the trajectories with a flip. Expanding $Q_\sigma(x)$ in Taylor's series and taking $\Delta t \to 0$ limit,  we get two coupled equations for $\sigma=\pm 1$,
\bea
[f(x_0) + v_0] \frac{d Q_+}{d x_0} - \gamma (Q_+(x_0) - Q_-(x_0)) &=& 0 \cr
[f(x_0) - v_0] \frac{d Q_-}{d x_0} + \gamma (Q_+(x_0) - Q_-(x_0)) &=& 0. \label{eq:QpQm_app}
\eea 
To solve these equations we need to determine the boundary conditions satisfied by $Q_\sigma(x_0)$. To this end, let us first consider the case where the particle starts from $x=-\ell_1,$ with $\sigma=-1$. In this case, the particle is expected to immediately exit the intermediate region through the left boundary. Similarly, if the particle starts at $x_0=\ell_1$ with $\sigma=1$, it is expected to exit through the right boundary. Thus, $Q_\pm(x_0)$ must satisfy the boundary conditions,
\bea
Q_+(\ell_1)=1, \quad \text{and}, \quad Q_-(-\ell_1)=0. \label{eq:BC}
\eea  
To solve the Eqs.~\eqref{eq:QpQm_app} it is convenient to recast these in terms of $Q(x_0) = Q_+(x_0)+Q_-(x_0)$ and $F(x_0) = Q_+ (x_0) - Q_-(x_0)$ as,
\bea
f(x_0)   \frac{d Q}{d x_0}  + v_0 \frac{d F}{d x_0} &=& 0  \cr 
f(x_0)   \frac{d F}{d x_0}  + v_0 \frac{d Q}{d x_0}   &=& 2 \gamma F(x_0). \label{eq:QF}
\eea
Multiplying the second equation by $f(x_0)$ and substituting the first equation,  we have a first order differential equation for $F(x_0)$,
\bea 
[f(x_0)^2 - v_0^2] \frac{d F}{d x_0} - 2 \gamma f(x_0) F(x_0)  =0.
\eea
This can be solved immediately to get,
\bea 
F(x_0) = A \exp \left[- 2 \gamma \int_{-\ell}^{x_0} \frac{du f(u)}{v_0^2 - f(u)^2}\right],
\eea 
where $A$ is an arbitrary constant. Using this solution in Eqs.~\eqref{eq:QF}, we obtain $Q_{\pm}(x_0)$,
\begin{align}
 Q_+(x_0) &=  \gamma \int_{-\ell}^{x_0} \frac{dy}{v_0 +f(y)} F(y) + B_1  \cr 
 Q_-(x_0) &=   \gamma \int_{-\ell}^{x_0} \frac{dy}{v_0 - f(y)} F(y) + B_2, 
\end{align}
where $B_1, B_2$ are arbitrary constants. It is easy to see that  $B_1, B_2$ and $A$ must be related by $B_1 - B-2 = A$, since $F (-\ell_1)= Q_+(-\ell_1) - Q_-(-\ell_1)$. Using this relation and the boundary conditions \eqref{eq:BC}, we can determine $Q_{\pm}(x_0)$ explicitly which are quoted in Eqs.~\eqref{eq:QpQm} in the main text.

\section{Computation of the barrier crossing time-scale} \label{ap:time-scale}

In this Appendix we provide the detailed steps involved in computing the barrier-crossing timescale $\tau$, starting from Eq.~\eqref{eq:tau_def}. From the definition of ${\cal N}$ in Eq.~\eqref{eq:norm} we can write,
\begin{align}
\tau & = 2 v_0 \int_0^{\ell} \frac{dx}{v_0^2 - f(x)^2} \exp \left[2 \gamma \int_0^x \frac{du f(u)}{v_0^2 - f(x)^2} \right].
\end{align}
Our goal is to investigate the behaviour of $\tau$ as $\epsilon = a_c - a \to 0$. To this end, we first note that, for small $\epsilon$, the dominant contribution to the above integral comes from near the maximum of $f(x)$, i.e, when $x$ is close to $m=a/\sqrt{3}$. Consequently, it is convenient to separate the integral into three regions: $0 \le x \le m-\Delta$, $ m-\Delta \le x \le m+\Delta$, and $ m+\Delta \le x \le \ell$ where $\Delta/m \lesssim 1$. For small $\epsilon = a_c -a$, the dominating contribution comes from the middle region, and we can write, 
\bea 
\tau &\simeq & 2v_0 \int_{m-\Delta}^{m+\Delta} \frac{dx}{v_0^2 - f(x)^2} \exp \left[2 \gamma \int_0^x \frac{du f(u)}{v_0^2 - f(x)^2} \right] \cr
&& + O(1). \label{eq:tau_int1}
\eea 
Let us first consider the behaviour of the integral inside the exponential, namely,
\bea
G(x) &=& 2 \gamma \int_0^x \frac{du f(u)}{v_0^2 - f(u)^2}.
\eea
Once again, for small values of $\epsilon$, this integral is dominated by the contribution from near  $u=m$, and we can write,
\bea 
G(x) &=&  2 \gamma \int_{m-\Delta}^x \frac{du f(u)}{v_0^2 - f(u)^2} + O(1). \label{eq:Gx_1}
\eea
Now, close to $u=m$, $f(u)$ has a quadratic form,
\bea
f(u) &\simeq & f(m) + \frac 12 (u-m)^2 f''(m) \cr  
 &=& v_0 \left(\frac{a}{a_c}\right)^3 - \sqrt{3}a (u-m)^2.
\eea
The above quadratic form leads to, to the second order in $(u-m)$,
\begin{align}
v_0^2 - f(u)^2 \simeq v_0^2 \left[1 - \left(\frac{a}{a_c} \right)^6 \right] + 2 \sqrt{3} v_0 \frac{a^4}{{a_c}^3} (u-m)^2. 
\end{align}
Using the definition of $a_c$ [see Eq.~\eqref{eq:ac_def}] this can be simplified to,
\begin{align}
v_0^2 - f(u)^2 \simeq \frac{4}{3}a^4 \left[\delta^2 + (u-m)^2 \right], \label{eq:v0_fu}
\end{align}
where, we have defined,
\begin{align}
\delta^2 = \frac{3 v_0^2 }{4 a^4} \left[1 - \left(\frac{a}{a_c} \right)^6 \right].
\end{align}
Now, using Taylor's expansion for small $\epsilon = a_c - a$, we have,
\begin{align}
\delta^2 = \frac 23 a_c \epsilon + O(\epsilon^2). \label{eq:del_eps}
\end{align}
Next, substituting \eqref{eq:v0_fu} in \eqref{eq:Gx_1}, we get, to the leading order in $\epsilon$,
\begin{align}
G(x) \simeq \frac{3 \gamma v_0}{2 a_c^4} \int_{m-\Delta}^x \frac{du}{\delta^2 + (u-m)^2} \label{eq:Gx_2}
\end{align}
where we have used the fact that $f(m)=v_0$ at $a=a_c$. This integral can be performed to obtain,
\begin{align}
G(x) \simeq \frac{\gamma} {\sqrt{3} a_c \delta} \left[\tan^{-1} \frac{x-m}{\delta} + \tan^{-1} \frac {\Delta}{\delta}\right]
\end{align}
For $\epsilon \to 0$, i.e., $\delta \to 0$, and $\Delta \sim O(1)$, $\tan^{-1} (\Delta/\delta) \to \pi/2$, while $\tan^{-1} ((x-m)/\delta) \to \text{sgn} (x-m) \pi/2$. Thus, 
\begin{align}
G(x) = \frac{\gamma \pi}{\sqrt{3} a_c \delta} \Theta(x-m) +O(1).
\end{align}
where $\Theta(z)$ denotes the Heaviside theta funtion. Substituting the above expression in Eq.~\eqref{eq:tau_int1}, we get,
\begin{align}
\tau = 2 v_0 A_1 \exp{\left(\frac{\pi \gamma}{\sqrt{3} a_c \delta}\right )} \int_m^{m+\Delta} \frac{dx}{v_0^2 - f^2(x)},
\end{align}
where $A_1$ is a constant which does not depend on $\epsilon$. Using Eq.~\eqref{eq:v0_fu} for the integrand, we get,
\begin{align}
\tau = \frac{3 v_0 A_1}{2 a_c^4 \delta} \exp{\left(\frac{\pi \gamma}{\sqrt{3} a_c \delta}\right )} \tan^{-1} \frac{\Delta}{\delta}.
\end{align}
In the limit $\delta \to 0$, $\tan^{-1} \frac{\Delta}{\delta} \to \pi/2$. Using Eq.~\eqref{eq:ac_def}, the above equation further simplifies to,
\begin{align}
\tau = A_1 \frac{\pi}{2 \sqrt{3} a_c \delta} \exp{\left(\frac{\pi \gamma}{\sqrt{3} a_c \delta}\right )}. 
\end{align}
Finally, the behavior of the escape time-scale near $a=a_c^-$ can be obtained using Eq.~\eqref{eq:del_eps}, 
\bea
\tau = \frac{A}{\sqrt{v_0(a_c -a)}} \exp \left[ \frac{\pi \gamma}{3^{3/4} \sqrt{v_0 (a_c -a)} } \right]
\eea
where $A$ is an undetermined factor, possibly depending on $\gamma, a$ and $v_0$. This is quoted in Eqs.~\eqref{eq:tau_a}-\eqref{eq:log_tau} in the main text.

\section{Numerical details}

We numerically simulate the Langevin dynamics Eq.~\eqref{eq:langevin}  using Euler discretization scheme with $dt=10^{-3}$. We have typically used $10^5$-$10^7$ realizations to measure the position distribution shown in Figs.~\ref{fig:Px_nonergo} and \ref{fig:a1}. To estimate the escape time, we measure the distribution $P(\tau_d)$  of the dwelling time $\tau_d$, which is defined as the time the RTP spends in a well until it crosses the barrier at $x=0$ for the first time.  The characteristic time estimated from the exponential tail of this distribution is shown in Fig.~\ref{fig:escape} as symbols.

\bibliographystyle{apsrev4-2}
\bibliography{ref}

\end{document}